\documentclass[fleqn,twoside]{article}
\usepackage{espcrc2}
\usepackage{graphicx}
\title{\bf Measurement of Inclusive Momentum Spectra and
Multiplicity Distributions of Charged Particles at $\sqrt{s} \sim$ 2 - 5 GeV}

\author{\small
\mbox{} \hskip 5cm (BES Collaboration) \\
\vspace{0.2cm}
J.~Z.~Bai$^1$, Y.~Ban$^{9}$,      J.~G.~Bian$^1$,
X.~Cai$^{1}$,       J.~F.~Chang$^1$,
H.~F.~Chen$^{16}$,  H.~S.~Chen$^1$,
Jie~Chen$^{8}$,    J.~C.~Chen$^1$,     Y.~B.~Chen$^1$
% ...
\\
\vspace{0.1cm}
$^1$ Institute of High Energy Physics, Beijing 100039, People's Republic of
     China\\
% ...
$^8$ Nankai University, Tianjin 300071, People's Republic of China\\
$^{9}$ Peking University, Beijing 100871, People's Republic of China\\
% ...
$^{16}$ University of Science and Technology of China, Hefei 230026,
        People's Republic of China\\
% ...
}

\begin{document}

\begin{abstract}
  {\bf{Abstract:}} Inclusive momentum spectra and multiplicity
  distributions of charged particles measured with the BESII detector
  at center of mass energies of 2.2, 2.6, 3.0, 3.2, 4.6 and 4.8 GeV
  are presented.  Values of the second binomial moment, $R_2$,
  obtained from the multiplicity distributions are reported. These
  results are compared with both experimental data from high energy
  $e^+e^-$, $ep$ and $p\bar{p}$ experiments and QCD calculations.

\vspace{1pc}
\end{abstract}

\maketitle
\section{Introduction}

Inclusive hadron production from $e^+e^-$ annihilations is very
valuable for testing Quantum Chromodynamics (QCD).
Perturbative QCD (pQCD) can give quantitative analytical predictions 
based on the Modified Leading Logarithmic Approximation
(MLLA)~\cite{mlla} under the assumption of Local Parton Hadron Duality
(LPHD)~\cite{lphd}.  Many experimental results have been reported from
high energy experiments~\cite{schmelling}
\cite{biebelxi}, which are in good agreement
with pQCD predictions.  However, hadron production has not been
studied well at low energies due to insufficient data, particularly at the
boundary region between pQCD and non-pQCD.
It is therefore very interesting to test pQCD in this region with low
energy $e^+e^-$ data.

Here we study the inclusive momentum spectrum, the hadron multiplicity
distribution, and the second binomial moment.  The inclusive momentum
spectrum is defined in terms of the variable $\xi = - \ln
(2p/\sqrt{s})$, where $p$ and $\sqrt{s}$ are the momentum of the
charged particles and center-of-mass (c.m.) energy respectively.  The
second binomial moment~\cite{r2nchthe}, a measure of the strength of
  hadron-hadron correlations and a sensitive probe for higher order
  QCD or non-perturbative effects~\cite{decamp}, is defined as $R_2
  = {\langle n_{ch}(n_{ch}-1) \rangle}/{\langle n_{ch} \rangle}^2$,
  where $n_{ch}$ is the charged particle multiplicity.  There has been
  a long standing discrepancy between the values of $R_2$ calculated
  to next to leading order (NLO) and those measured in $e^+e^-$
  experiments \cite{schmelling}. 


In this paper, we report measurements of the inclusive momentum
spectra and multiplicity distributions of charged particles obtained with
the upgraded Beijing Spectrometer (BESII) at the Beijing
Electron-Positron Collider (BEPC) with center-of-mass energies of 2.2,
2.6, 3.0, 3.2, 4.6 and 4.8 GeV.  Results are compared with pQCD
calculations, as well as those from high energy $e^+e^-$, $ep$ and
$p\bar{p}$ experiments.

\section{Detector and trigger}

The measurements were done using the data collected for the $R$
scan~\cite{besr_1,besr_2} with BESII, a conventional solenoidal magnet
detector that is described in detail in Ref.~\cite{bes}. A vertex chamber
comprising 12 tracking layers surrounds a
beryllium beam pipe and provides input to the trigger system,
as well as coordinate information.  The primary
tracking device is the cylindrical main drift chamber
(MDC). It has 40 layers of sense wires and yields
precise measurements of charged particle trajectories;
it also provides $dE/dx$ information which is used for
charged particle identification. Outside the MDC, there
is a barrel time-of-flight system (BTOF) consisting of an
array of 48 plastic scintillator counters that are read out
at both ends by fine-mesh photomultiplier tubes.
Electron and photon
showers are detected in a sampling barrel shower
counter (BSC) that covers $80 \%$ of the total solid angle.
It consists of 24 layers of
self-quenching streamer tubes interspersed with lead;
each layer has 560 tubes. The outermost component of BESII is
a muon identification system consisting of three double layers
of proportional tubes interspersed in the iron flux return of the
magnet.

The triggers and the measurement of the trigger efficiencies for
Bhabha, dimuon and hadronic events are the same as those
described in Ref.~\cite{besr_1}. The efficiencies are measured by comparing
the responses to different trigger requirements in special runs
taken at the peak of the $J/\psi$ resonance, and are determined to 
be 99.96\%, 99.33\% and 99.76\% for Bhabha, dimuon, and hadronic events,
respectively. Their uncertainties are about $\pm 0.5\%$.

\section{Hadronic event selection and background subtraction}

The hadronic event samples are almost the same as those used for the
measurement of $R$ values at these energies. 
Sources of background are cosmic rays,
pair produced leptons, two-photon processes, and single-beam associated
processes. First, clear Bhabha events are rejected from the sample.
Next, hadronic events are selected.  Special attention is paid to
two-prong events, where cosmic ray and lepton pair backgrounds are
especially severe, and additional requirements are imposed to provide
extra background rejection~\cite{besr_1}. 

The hadron selection proceeds by first selecting charged tracks and then
selecting events.
For the track-level selection, we require the track polar angle satisfy
$|\cos \theta| < 0.80$, the track must not be identified as an electron
or a muon, and the distances of closest approach to the beam in the
transverse plane and along the beam axis should be less than 2.0 and 18
cm, respectively.  The following criteria are also used to define
``good'' charged tracks:
\begin{itemize}
\item[(i)]  $p < E_{beam} + (5 \times \sigma_{p})$, where $p$ and
      $E_{beam}$ are the momenta of the track and the beam energy,
      respectively, and $\sigma_p$ is the momentum resolution
      for charged tracks at $p=E_{beam}$;
\item[(ii)] $E < 0.6 E_{beam}$, where $E$ is the energy in the BSC
      that is associated with the track;
\item[(iii)] $2 < t < t_{p} + (5 \times \sigma_{t})$ (in ns), where
      $t$ and $t_{p}$ are the time-of-flight for the track
      and a nominal time-of-flight calculated for the track
      assuming a proton hypothesis, and
      $\sigma_{t}$ is the BTOF time resolution.
\item[(iv)]  $p_t \geq$ 0.08GeV, where $p_t$ is the transverse
      momentum.
\end{itemize}

The event-level selection requires at least two charged tracks with at
least one ``good'' track satisfying the above requirements, the
total deposited energy in the BSC should be greater than
$0.28 E_{beam}$, and the tracks selected should not all point
along either the $+z$ or $-z$ direction.
These criteria help reject background. For two-prong events, 
cosmic-ray and
lepton pair events are removed by requiring that tracks should
not be back-to-back and that there should be at least two isolated
neutral clusters with $E > 100$ MeV and with the differences in
azimuthal angle
with charged tracks more than $15^{\circ}$. This last requirement
rejects radiative Bhabha events.
The above selection procedures
remove most backgrounds. Backgrounds from two-photon processes
are negligible after hadron selection.

Residual background contributions from Bhabha processes and
tau pair production are subtracted using a Monte Carlo analysis, e.g. for
Bhabha events, the number of Bhabhas in the hadronic events is
\begin{equation}
N_{bb} = \frac{N_{bb}^{pass}}{N_{bb}^{gen}}
         {\mathcal{L}} \cdot \sigma_{bb},
\end{equation}
where $N_{bb}^{pass}$ is the number of Bhabha events surviving
hadronic event selection, $N_{bb}^{gen}$ is the total number of
generated Bhabha events, $\mathcal{L}$ is the integrated luminosity at
each energy point,
and $\sigma_{bb}$ is the total Bhabha cross-section calculated by
the Monte Carlo program.  
A detailed comparison between data and Monte Carlo data 
shows that the maximum discrepancy in any momentum bin is 
10\%; hence the 
error due to the Monte Carlo simulation of Bhabha events and tau
decays
is less than 1\% in any bin, 
far less than the 5-10\% statistical uncertainty.

The beam-associated background is treated by fitting the distribution of
event vertices along the beam direction with a Gaussian for real
hadronic events and a second order polynomial for the background, as
described in Refs.~\cite{besr_2}\cite{beamgas}.  
Table \ref{hadron} summarizes the background contributions to the hadron
samples from Bhabhas, tau decays, and beam-associated events. 
The typical background is a few percent below $\tau \tau$ threshold
and about $15 \%$ above. 

We do not subtract the contributions from the $\psi(4040)$,
$\psi(4160)$, or the $\psi(4415)$ resonances to the high energy
points. Using the resonance parameters from the Particle Data
Group~\cite{pdg2002} and varying them within errors, the contribution
to the 4.6 GeV point is expected to be below 2.2 \%, while that at the
4.8 GeV point is less than 1 \%.

\begin{table}[htbp]
\begin{center}
\caption{Background contribution to the hadron sample.}
\begin{tabular}{ccccc}\hline
$E_{cm}$  & $N_{had}$ & Beam-assoc.  & $e^+e^-$    &
$\tau$  pair    \\
(GeV)        &         & (\%)    &(\%)   & (\%)    \\
\hline
2.2  & 1410 & 3.82 & 0.61  & ---  \\
2.6  & 4968 & 3.72 & 0.48  & ---  \\
3.0  & 2030 & 3.01 & 0.56  & ---  \\
3.2  & 1828 & 4.53 & 0.35  & ---  \\
4.6  & 1315 & 6.86 & 0.18  & 6.98 \\
4.8  & 1282 & 8.72 & 0.14  & 6.34 \\ \hline
\end{tabular}
\label{hadron}
\end{center}
\end{table}

\section{Hadronic event generator and initial state radiative
correction}

\begin{figure}[htb]
\includegraphics[width=17.5pc]{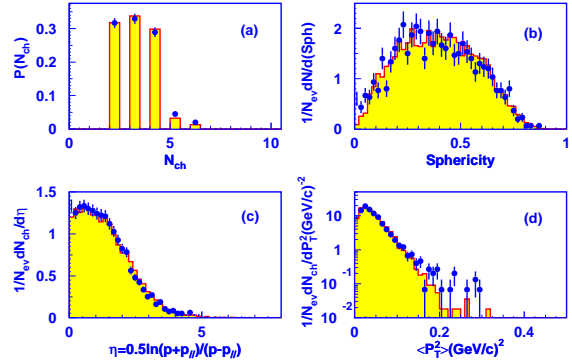}
\caption{Comparison between
data (dots with error bars) and the LUARLW Monte Carlo (histogram) at
2.2 GeV: (a) Multiplicity; (b) Sphericity; (c) Rapidity; and (d)
Transverse momentum. }
\label{comdtal}
\end{figure}

Hadronic events are simulated by the JETSET7.4 \cite{jetset} and
LUARLW~\cite{huarlw} Monte Carlo programs.  LUARLW is used for
energies at 3 GeV or lower.  JETSET with tuned parameters is used in
the region from 3 to 4 GeV, and JETSET with default parameters is used
for 4 GeV and higher energy points \cite{besr_2}.  The detector
response is based on EGS for electromagnetic interactions, while for
hadronic interactions, parameterizations are used.  Above 3.77 GeV,
the production of $D,~D^*, ~D_s,$ and $D_s^*$ is included in the
generator according to the Eichten Model~\cite{eichten}.  A Monte
Carlo event generator has been developed to handle decays of the
resonances in the ``radiative return'' processes $e^+e^- \rightarrow
\gamma J/\psi$ or $\gamma \psi(2S)$ \cite{chenjc}.  Four
distributions comparing experimental data and Monte Carlo
data at 2.2~GeV are shown in Fig.  \ref{comdtal}. The data and Monte
Carlo distributions agree reasonably well.

Initial state radiation effects are simulated by a Monte Carlo program
which is based on the scheme described in Ref.~\cite{berends}. The error
on the effective radiative correction factor $(1+\delta_{obs})$ is 
estimated to be less than $3\%$ by comparing different
formalisms \cite{bonneau}\cite{kureav} and using different cuts.

\section{$\xi$ spectrum}

The charged particle inclusive hadron momentum spectrum can be expressed as a function of 
$\xi = - \ln(2p/\sqrt{s})$. A purely analytical approach giving
quantitative predictions for $\xi$ is a QCD calculation using MLLA~\cite{mlla}
under the assumption of LPHD~\cite{lphd},
expressed as
\begin{eqnarray}
\frac{1}{\sigma_{had}} \frac{d \sigma}{d \xi} = 2K_{LPHD} \times
        f_{MLLA} (\xi,\Lambda_{eff}, N_c,n_f)
\label{express}
\end{eqnarray}
where $K_{LPHD}$ is an overall normalization factor
describing hadronization, $f_{MLLA}$ is a complex function of $\xi$
and the effective scale parameter $\Lambda_{eff}$,
$N_c$ is the color factor, and $n_f$ is the number of active
quarks.
Eq.~\ref{express} is only valid in the range $0 \leq \xi \leq \ln
(0.5\sqrt{s} / \Lambda_{eff})$.

The $\xi$ spectrum at each 
center-of-mass energy is obtained
using the following: 
\begin{eqnarray}
\frac{1}{\sigma_{had}} \frac{d \sigma}{d\xi_i} =
C(\xi_i) \frac{1}{N_{had}} \frac{N_{obs}(\xi_i)}{\Delta \xi_i},
\end{eqnarray}
where, $N_{had}$ is the total number of hadronic events observed,
$N_{obs}(\xi_i)$ is the number of charged tracks in bin-$i$, $\Delta
\xi_i$ is the bin width chosen commensurate with the detector
resolution to avoid significant migration between bins, and $C(\xi)$
is a multi-source correction factor including effects from the
detector, ISR, and hadronic event selection.  This correction is
determined by two Monte Carlo samples: sample I (hadron level)
includes neither initial state radiation (ISR) nor detector simulation
and sample II (detector level) includes both ISR and detector
simulation . Events of sample II are reconstructed in the same way as
for data and subjected to the same selection criteria.  Distributions
at the hadron level are only from particles with a lifetime greater
than $3 \times 10^{-10}$ s, which are considered as stable
\cite{stable}.

The correction factors $C(\xi)$
for each bin are given by
\begin{eqnarray}
C(\xi_i)=(\frac{N_{gen}(\xi_i)}{N_{gen}^{total}}) /
       (\frac{N_{det}(\xi_i)}{N_{det}^{total}})
\label{corrfactor}
\end{eqnarray}
where $N_{gen}^{total}$ is the total number of
generated Monte Carlo events for sample I, $N_{det}^{total}$
is the number of events which pass the hadron selection in II, and
$N_{gen}(\xi_i)$ and $N_{det}(\xi_i)$ are the number of entries in the
ith bin, respectively.  These correction factors are determined with
the LUARLW and JETSET Monte Carlo simulations at 3 GeV, where
both Monte Carlo generators are valid. Their differences are included
as systematic errors for all center-of-mass
energies. Other systematic errors include variations due to selection
criteria changes, etc.

\begin{figure}[htbp]
\includegraphics[width=17.5pc]{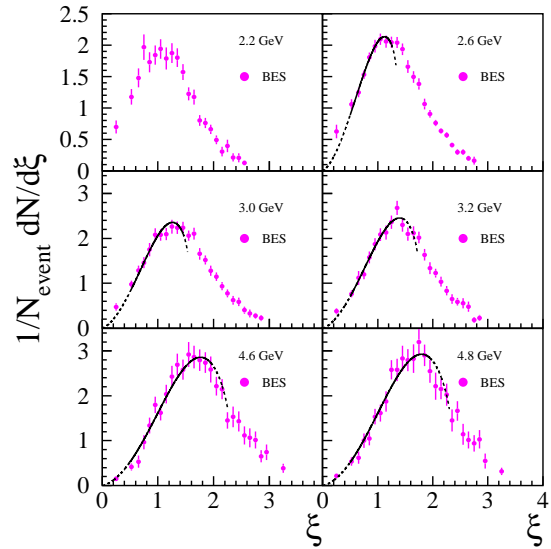}
\caption{Measured $\xi$ spectra (solid dots) at 2.2, 2.6,
3.0, 3.2, 4.6 and 4.8 GeV.  The solid curves show the data region fitted
with the limiting spectrum. The dotted line is an
extrapolation of the fitted result.} \label{comxi}
\end{figure}

The measured $\xi$ spectra at the six energies between
2.2 and 4.8 GeV are shown in Fig. \ref{comxi}. 
For each point, 
we add the statistical and systematic errors 
in quadrature.
These spectra are fitted with Eq. \ref{express} in the range
$0.5 \leq\xi\leq \ln (0.5\sqrt{s}/p_0)$, where 
$p_0 \sim 0.35$ GeV.
The solid lines in
the plots indicate the actual fitting range while the
dashed line (outside the fitted range) indicate the QCD calculations 
using the fitted parameters, $K_{LPHD}$ and $\Lambda_{eff}$,
which are shown in Table \ref{klphdeff}.  At 2.2 GeV the fit range is
very restricted and little of the peak region remains for comparison
with MLLA; therefore this point is excluded.
The fitted values of
$\Lambda_{eff}$ increase with decreasing $E_{cm}$.
The differences from varying 
fitting ranges are included as systematic errors. 

\begin{small}
\begin{table}[htbp]
\begin{center}
\caption{The fitted $\Lambda_{eff}$ and $K_{LPHD}$ values. The first
uncertainty is statistical and second systematic.}
\begin{tabular}{|c|c|c|c|c|}\hline
Ecm   & $\Lambda_{eff}$ (MeV) & $K_{LPHD}$  \\ \hline
2.6   & 342 $\pm$ 7 $\pm$ 23 & 1.523 $\pm$ 0.018 $\pm$ 0.023 \\
3.0   & 325 $\pm$ 9 $\pm$ 25 & 1.573 $\pm$ 0.027 $\pm$ 0.026 \\
3.2   & 286 $\pm$ 17 $\pm$ 37 & 1.532 $\pm$ 0.028 $\pm$ 0.052 \\
4.6   & 239 $\pm$ 14 $\pm$ 32 & 1.472 $\pm$ 0.029 $\pm$ 0.039 \\
4.8   & 238 $\pm$ 15 $\pm$ 32 & 1.482 $\pm$ 0.029 $\pm$ 0.038 \\
\hline
\end{tabular}
\label{klphdeff}
\end{center}
\end{table}
\end{small}

Eq. \ref{express} has a maximum in each $\xi$ spectrum as shown in
Fig.~\ref{comxi}. Using the fitted parameters $\Lambda_{eff}$ from
Table \ref{klphdeff}, the peak positions, $\xi^{\star}$, can be
determined directly from the limiting spectrum (Eq. \ref{express}).

The peak positions can also be obtained 
by fitting a Gaussian or distorted 
Gaussian~\cite{disg} over a limited range of 
$1 / N_{event} \times dN / d\xi \geq 0.6 (1 / N_{event}
\times dN / d\xi)_{max}$. 
We choose the Gaussian fit since the
distorted Gaussian gives consistent results but with larger
errors due to the larger number of free parameters.
The changes found when the fitting range is varied are included as 
systematic uncertainties. Table \ref{peakfit} lists the fitted
results. The results of the two methods (pQCD prediction of
Eq. \ref{express}
and fitted
Gaussian form) are consistent at higher energies, and the differences 
between the two methods are also 
included in the systematic uncertainties.


\begin{table}[htbp]
\begin{center}
\caption{Peak positions $\xi^{\star}$ of the $\xi$ spectrum found by
  fitting with a Gaussian. The
first uncertainties are statistical and the second systematic.}
\begin{tabular}{|c|c|}\hline
Ecm (GeV) &   $\xi^{\star}$  \\ \hline 
2.2  & 1.063 $\pm$ 0.017 $\pm$ 0.032 \\\hline
2.6  & 1.189 $\pm$ 0.012 $\pm$ 0.076 \\\hline
3.0  & 1.304 $\pm$ 0.017 $\pm$ 0.049 \\\hline
3.2  & 1.361 $\pm$ 0.021 $\pm$ 0.057 \\\hline
4.6  & 1.646 $\pm$ 0.025 $\pm$ 0.134 \\\hline
4.8  & 1.633 $\pm$ 0.037 $\pm$ 0.193 \\\hline
\end{tabular}
\label{peakfit}
\end{center}
\end{table}

\begin{figure}[htbp]
\includegraphics[width=17.5pc]{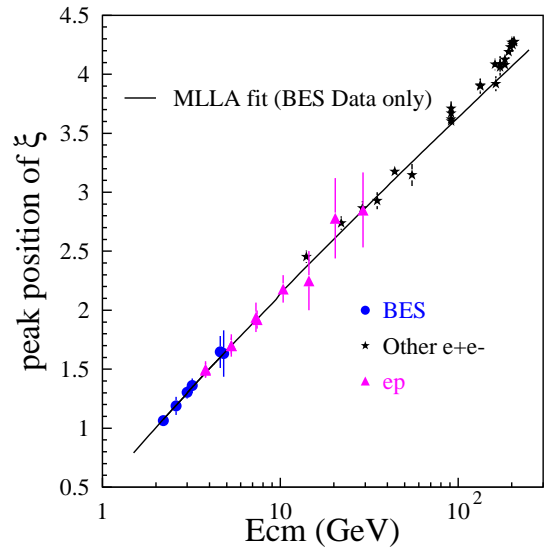}
\caption{Energy dependence of the peak position
$\xi^{\star}$ from Table~\ref{peakfit}. The curve is obtained fitting
with only the BES data by Eq.~\ref{peakstar}, but results from
other $e^+ e^-$ \cite{schmelling,biebelxi} and $e p$ \cite{epxi}
experiments are also plotted.} \label{peakbes}
\end{figure}

The measured peak positions from this work are plotted in
Fig.~\ref{peakbes}, together with those of high energy $e^+e^-$ and
$ep$ data.  MLLA/LPHD predicts the energy dependence of the peak
position, $\xi^{\star}$, as~\cite{mlla}
\begin{equation}
\xi^{\star} = 0.5 Y + \sqrt{c Y} - c,
\label{peakstar}
\end{equation}
where $Y = \ln (0.5\sqrt{s}/\Lambda_{eff})$ and $c$ is 0.2915 (0.3190)
for three (four) active flavors. By fitting our data to
Eq.~\ref{peakstar} , we obtain $\Lambda_{eff} = 262 \pm 9$ MeV. This
value is consistent with the results from OPAL~\cite{opalxi2},
ZEUS~\cite{zeusxi} and CDF~\cite{cdfxi}, which are $263 \pm 4$, $251
\pm 14 (ep)$, and $256 \pm 13 (p\bar{p})$ MeV, respectively. However,
our value disagrees with L3~\cite{l3xi} and the result of
Ref.~\cite{biebelxi}, which are $200 \pm 3$ and $232\pm 3$,
respectively. Figure~\ref{peakbes} shows that $\xi^{\star}$ is
approximately linear in $\ln \sqrt{s}$. A straight line fit of the BES
results to $\xi^{\star}$ as a function of $\sqrt{s}$ gives a gradient
of $0.779 \pm 0.122$. The gradients reported by OPAL~\cite{opalxi},
ZEUS~\cite{epxi} and H1~\cite{h1nch} are $0.637 \pm 0.016$, $0.650 \pm
0.077$, and $0.75 \pm 0.05$, respectively. Our result is somewhat
larger.


\section{Momentum spectrum}

\begin{figure}[htbp]
\includegraphics[width=17.5pc]{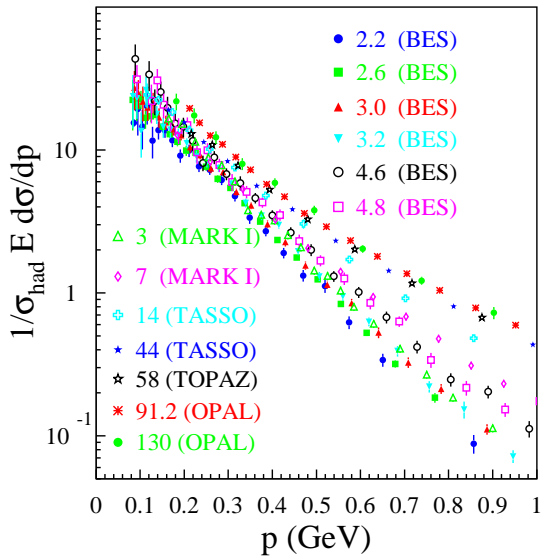}
\caption{Charged particle momentum spectra for center of
  mass energies from 2.2 to 130 GeV. Also shown are MARKI
  \cite{MARKI82}, TASSO \cite{tasso90}, TOPAZ \cite{itoh}, and OPAL
  \cite{alexander,opalxi} results.}
\label{dnall}
\end{figure}

In order to study the behavior of low momentum particles, MLLA/LPHD
introduces a Lorentz invariant variable $\frac{1}{\sigma_{tot}} E
\frac{d\sigma}{dp}$, where $E^2 = P^2 + Q_0^2$ and $Q_0 = 0.27$ GeV
\cite{q0momentum}.  The six BES Lorentz invariant, charged particle
momentum spectra are plotted in Fig.~\ref{dnall}, together with those
measured by other experiments at higher energy, up to 130 GeV. The
tendency of the spectra to approximately converge at lower momenta
is cited as evidence  
that hadron production at very small momentum $p
\leq 0.1$ GeV is approximately energy independent. \cite{abreu}. This
behavior has been explained in Ref. \cite{q0momentum} as the coherent
emission of low energy (i.e.  long wavelength) gluons by the total
color current.

\section{Charged particle multiplicity and the second binomial moment}

According to a NLO QCD calculation, the second binomial moment $R_2$
is given by \cite{r2nchthe}

\begin{eqnarray}
R_2    = \frac{11}{8}\bigg[1 - c \sqrt{\alpha_s(\sqrt{s})}\bigg]
\label{r2exp}
\end{eqnarray}

\noindent with $c = 0.55 (0.56)$ for five (three) active flavors.

The charged
multiplicity distributions are obtained using a method which was first
employed by TASSO \cite{tasso} and modified by H1 \cite{h1nch}.
Three steps are used to correct the measured unnormalized charged
multiplicity distribution $N_{det}^{exp}(i)$, where $i$ is the
number of observed tracks. The first step is to correct for
detector effects and selection criteria. It is performed using Monte Carlo
sample II with its hadron level information. For each event, the
number of observed tracks $j$ with distribution $N_{det}^{MC}(j)$
is compared to the number of generated $k$ tracks with
distribution $N_{gen}^{MC}(k)$. Denoting $N_{jk}$ as the
number of events generated with $k$ tracks when $j$ tracks are
observed and $N_j$ as the number of observed events with $j$ tracks,
the correction matrix $M(k,j)$ is expressed as

\begin{equation}
M(k,j) = N_{jk} / {N_j}.
\end{equation}

\noindent This matrix relates $N_{det}^{MC}(j)$ to
$N_{gen}^{MC}(k)$ by

\begin{eqnarray}
N_{gen}^{MC}(k) = \sum_{j} M(k,j) \cdot N_{det}^{MC}(j)
\end{eqnarray}

The second step is to correct for the presence of QED initial
state radiation which results in a reduction of the nominal
c.m. energy and thus changes the charged multiplicity. Therefore
another set of correction factors is calculated according to
\begin{equation}
C_F(k) = \rho_{NR}(k) / \rho_{gen}(k),
\end{equation}
where $\rho_{NR}(k)$ is the normalized multiplicity
distribution for events generated at fixed c.m. energy without
detector simulation and $\rho_{gen}(k)$ is the distribution
$N_{gen}^{MC}(k)$ after normalization. Finally, the corrected
multiplicity distribution $N_{exp}^{corr}$ reads
\begin{eqnarray}
N_{exp}^{corr}(k) = C_F(k) \sum_{i} M(k,i)
                               \cdot N_{det}^{exp}(i).
\end{eqnarray}

The final step is done to reduce bias that might be introduced by the Monte
Carlo generator used \cite{h1nch}.  
An iterative procedure is used where the predicted
multiplicity at the hadron level is reweighted using the previously
unfolded multiplicity.  This is repeated until convergence is reached.

\begin{figure}[htbp]
\includegraphics[width=17.5pc]{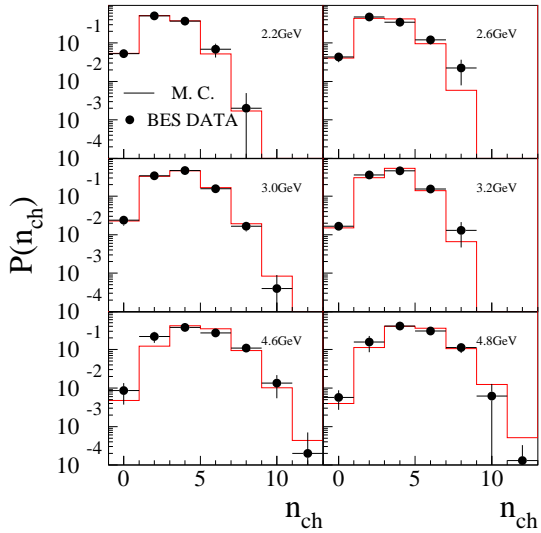}
\caption{Multiplicity distributions at energies
of 2.2, 2.6, 3.0, 3.2, 4.6 and 4.8 GeV. Solid dots are from this
work; histograms are predicted by Monte Carlo.} \label{comnch}
\end{figure}

\begin{figure}[htbp]
\includegraphics[width=17.5pc]{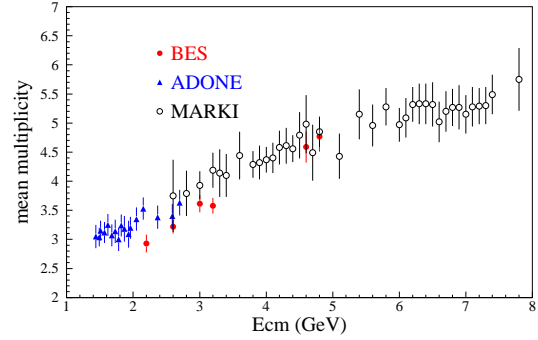}
\caption{Energy dependence of mean multiplicity for
energies below 8 GeV.  Also shown are results of ADONE \cite{ADONE}
and MARKI \cite{MARKI82}.} \label{meannch}
\end{figure}

The measured multiplicity distributions at different energies are
shown in Fig. \ref{comnch} along with the Monte Carlo predictions.
Figure \ref{meannch} shows the energy dependence of the mean
multiplicity defined as $\langle n_{ch} \rangle =
\sum\limits_{n_{ch}=0}^{\infty} n_{ch} P(n_{ch})$, where $P(n_{ch})$ is
the probability to have $n_{ch}$ tracks.

\begin{figure}[htbp]
\includegraphics[width=17.5pc,clip]{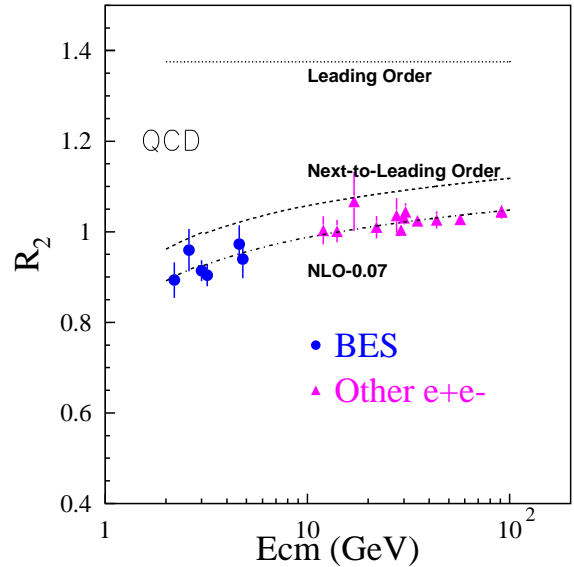}
\caption{Energy dependence of the second binomial
moment $R_2$. For other $e^+ e^-$ data, see References \cite{schmelling,biebelxi}.} \label{r2bes}
\end{figure}

Based on the measured multiplicity distributions, we obtain the second
binomial moments $R_2$, which are displayed in Fig.~\ref{r2bes},
together with both NLO calculations and published data at higher
energies up to 100 GeV from other $e^+e^-$ experiments.
Our measured $R_2$ values, though with large errors, are consistent with
other measurements at higher energies.
The $R_2$ value predicted by leading order QCD
is significantly higher than the measured data, and while the NLO
calculation comes closer to the data, the remaining disagreement
of about $\sim 0.07$ in $R_2$ may indicate
that $R_2$ is a sensitive probe for higher order QCD or
non-perturbative effects.

\section{Summary}
Using BESII $R$ scan data, we have measured the inclusive momentum
spectra, multiplicity distributions, and the
second binomial moments at $\sqrt{s} \sim$ 2 - 5 GeV in the continuum.
These results are compared with data from other
experiments at higher energies, and with QCD model calculations.

The parameters $K_{LPHD}$ and $\Lambda_{eff}$ obtained by fitting
the limiting spectra are compatible with those of the ZEUS, CDF, and OPAL
experiments. However, the deviations become larger at energies
below $3$ GeV. 


The second binomial moment $R_2$ determined by the BES experiment, though with
large errors, are lower than what is predicted to NLO.
This is consistent with the long standing discrepancy between the NLO
calculation and high energy data from $e^+e^-$ experiments \cite{schmelling}.

To further reduce the uncertainties of our measurements and test
QCD model calculations to a precision of a few percent, higher
statistics data collected with a better detector are needed. This
could be one of the interesting physics goals of the CLEOc and
BESIII projects~\cite{cleoc,bes3}.

\section{Acknowledgment}
We thank the IHEP Accelerator, Computing and BES/BEPC Support
divisions. We are especially grateful to Prof. W. Ochs for his help
on MLLA/LPHD theory and providing the experimental $e^+e^-$ data at high
energy in Fig. 4.

This work is supported in part by
the National Natural Science Foundation
of China under contracts No. 19991480, No. 19805009 and
No. 19825116; the Chinese Academy of Sciences under contract
No. KJ 95T-03, No. E-01 (IHEP) and No. X-45 (IHEP); and
by the Department of Energy under Contract No.
DE-FG03-93ER40788 (Colorado State University),
DE-AC03-76SF00515 (SLAC), DE-FG03-94ER40833 (U Hawaii),
DE-FG03-95ER40925 (UT Dallas).

\end{document}